\documentstyle[twoside,fleqn,espcrc2]{article}
\def\simlt{\stackrel{<}{{}_\sim}}
\def\simgt{\stackrel{>}{{}_\sim}}
\def\NPB#1#2#3{{\it Nucl.~Phys.} {\bf{B#1}} (19#2) #3}
\def\PLB#1#2#3{{\it Phys.~Lett.} {\bf{B#1}} (19#2) #3}
\def\PRD#1#2#3{{\it Phys.~Rev.} {\bf{D#1}} (19#2) #3}

\def\AP#1#2#3{{\it Ann.~Phys.} {\bf#1} (19#2) #3}

\newcommand{\AmS}{{\protect\the\textfont2
  A\kern-.1667em\lower.5ex\hbox{M}\kern-.125emS}}

\begin{document}
\begin{titlepage}\phantom{.}\vskip -3cm 
\begin{flushright}
\large
CERN--TH/97-232\\[0.1mm]
CPTH--PC555.0897\\[.1mm]
IEM--FT--164/97\\[.1mm]
hep-th/9709023
\end{flushright}\vskip -1cm
\begin{center}\Large\bf Supersymmetry breaking in M-theory$^*$\\[1cm]
\Large I. Antoniadis,$^{a,b}$ M. Quir\'os$^c$\\[1cm]
\large\sl
$^a$TH-Division, CERN, CH-1211 Geneva 23, Switzerland\\[1mm] \large\sl
$^b$Centre de Physique Th{\'e}orique, Ecole Polytechnique, F-91128 
Palaiseau cedex, France\\[1mm] \large\sl
$^c$Instituto de Estructura de la Materia, CSIC, Serrano 123, 
28006-Madrid, Spain\\[1cm]
\bf Abstract
\end{center}
{\large We describe the breaking of supersymmetry in M-theory
by coordinate dependent (Scherk-Schwarz) compactification of the
eleventh dimension. Supersymmetry is spontaneously broken in the
gravitational and moduli sector and communicated to the
observable sector, living at the end-point of the semicircle, by
radiative gravitational interactions. This mechanism shares the generic features
of non-perturbative supersymmetry breaking by gaugino
condensation, in the presence of a constant antisymmetric field
strength, in the weakly coupled regime of the heterotic string,
which suggests that both mechanisms could be related by duality. In
particular an analysis of supersymmetric transformations 
in the infinite-radius limit reveals the presence of a discontinuity 
in the spinorial parameter, which coincides with the result found in 
the presence of gaugino condensation, while the condensate is identified 
with the quantized parameter entering the boundary conditions. }
\\[1cm]

\noindent
Based on invited talks delivered at:
{\it 5th International Conference on Supersymmetries in Physics},
{\bf SUSY 97}, May 27-31, 1997, University of Pennsylvania, Philadelphia, 
PA, USA;
{\it International Europhysics Conference on High Energy Physics}, 
{\bf HEP 97}, 19-26 August, Jerusalem, Israel.\\[.1cm]
\hrule\vskip 2mm\noindent
$^{*}$Research supported
in part by IN2P3-CICYT under contract Pth 96-3, in part by CICYT contract 
AEN95-0195, and in part by the EEC under the TMR contract ERBFMRX-CT96-0090.
\begin{flushleft}
\large
CERN--TH/97-232\\[3mm]
September 1997
\end{flushleft}
\end{titlepage}   

\hyphenation{author another created financial paper re-commend-ed}

\title{Supersymmetry breaking in M-theory\thanks{Research supported
in part by IN2P3-CICYT under contract Pth 96-3, in part by CICYT contract 
AEN95-0195, and in part by the EEC under the TMR contract ERBFMRX-CT96-0090.}}
\author{I. Antoniadis\address{TH-Division, CERN, CH-1211 Geneva 23, 
Switzerland}\thanks{On leave from: 
Centre de Physique Th\'eorique, Ecole Polytechnique, France}
      and 
        M. Quir\'os\address{Instituto de Estructura de la Materia (CSIC),
        Serrano 123, E-28006 Madrid, Spain}  }

\begin{abstract}
We describe the breaking of supersymmetry in M-theory
by coordinate dependent (Scherk-Schwarz) compactification of the
eleventh dimension. Supersymmetry is spontaneously broken in the
gravitational and moduli sector and communicated to the
observable sector, living at the end-point of the semicircle, by
radiative gravitational interactions. This mechanism shares the generic features
of non-perturbative supersymmetry breaking by gaugino
condensation, in the presence of a constant antisymmetric field
strength, in the weakly coupled regime of the heterotic string,
which suggests that both mechanisms could be related by duality. In
particular an analysis of supersymmetric transformations 
in the infinite-radius limit reveals the presence of a discontinuity 
in the spinorial parameter, which coincides with the result found in 
the presence of gaugino condensation, while the condensate is identified 
with the quantized parameter entering the boundary conditions. 
\end{abstract}

\maketitle

\section{Introduction}
Models of particle physics derived from the 10-dimensional (10D) 
$E_8\times E_8$ heterotic string, compactified on an
appropriate 6D internal manifold, are the most attractive candidates for 
describing the observed low-energy world. In particular, compactification 
on a Calabi-Yau (CY) manifold leads to a 4D $N=1$ supersymmetric theory
that can accommodate the gauge group and matter content of the standard model.
One difficulty is the mismatch between the gauge coupling unification 
scale, $M_G\sim 10^{16}$ GeV, and the heterotic string scale $M_H$, which is
determined in terms of the Planck mass, $M_p\sim 10^{19}$ GeV, as
\begin{equation}
M_H=(\alpha_G/8)^{1/2}M_p\sim 10^{18}\ {\rm GeV}, 
\label{stringscale}
\end{equation}
where $\alpha_G\sim 1/25$ is the unification gauge coupling. 
However, the perturbative relation $M_G=M_H$ 
does not hold if the compactification scale 
$V^{-\frac{1}{6}}$ [compactification volume$\equiv(2\pi)^6 V$] $\ll M_H$,
in which case the 10D theory is strongly interacting,
\begin{equation}
\lambda_H=2(\alpha_G V)^{1/2} M_H^3 \gg 1
\label{lambda}
\end{equation}
and the scales mismatch can be given an interesting solution. 

\section{Strongly coupled $E_8\times E_8$ heterotic string}
The strong coupling limit of the heterotic $E_8\times E_8$ superstring
compactified on a $CY$ manifold is believed to be described by the 
eleven-dimensional M-theory compactified on $CY\times S^1/Z_2$, where the
semicircle has a radius $\rho$~\cite{hw}. The
relations between the eleven- and ten-dimensional parameters are
\cite{w}:
\begin{equation}
M_{11}=M_H \left( {\sqrt{2}\over\lambda_H}\right)^{1/3}\quad
\rho^{-1}={1\over\sqrt{2}\lambda_H}M_H\ ,
\label{HM}
\end{equation}
where we have defined the eleven-dimensional scale $M_{11}=2\pi
(4\pi\kappa_{11}^2)^{-1/9}$ \cite{ckm} and $\kappa_{11}$ is the 11D
gravitational
coupling. When the ten-dimensional heterotic coupling is
large ($\lambda_H\gg 1$), the radius of the semi-circle is large and M-theory is
weakly coupled on the world-volume. 

Using eqs.~(\ref{stringscale}) and (\ref{lambda}), one can express
$M_{11}$ and $\rho$ in
terms of the four-dimensional parameters:
\begin{equation}
M_{11}=(2\alpha_G V)^{-1/6}\quad
\rho^{-1}= {4\over\alpha_G}M_{11}^{3}M_p^{-2}\ .
\label{M11}
\end{equation}
In this regime, the value of the unification scale $M_G$ becomes 
$\sim M_{11}$, which is lower than $M_H$ (and can be fixed to the desired value
because of $V$), while the radius $\rho$ 
of the semicircle is at an intermediate scale
$\rho^{-1}\sim 10^{12}$ GeV, and for isotropic CY the compactification scale 
$V^{-1/6}$ is of the order of $M_{11}$\cite{m0}. Fortunately, 
this is inside the region of validity of M-theory, 
$\rho M_{11}\gg 1$ and $(2\pi)^6 V\kappa_{11}^{-4/3}\gg 1$.
As a result, the effective theory above the intermediate scale behaves as
5-dimensional, but only in the gravitational and moduli sector; the gauge
sectors coming from $E_8\times E_8$ live at the 4D boundaries 
of the semicircle.

\section{Compactification of M-theory on $CY\times S^1/{\bf Z}_2$}

Here, we review the main properties of M-theory compactification in four
dimensions on a seven-dimensional internal space, which is the product of a 
Calabi--Yau manifold with the semicircle $S^1/Z_2$. Proceeding in 
two steps, we will first consider the five-dimensional theory on a Calabi--Yau
threefold with Hodge numbers $h_{(1,1)}$ and $h_{(2,1)}$ leading to $N=1$ 5D
space-time supersymmetry~\cite{ccaf}. 
In addition to the gravitational multiplet, 
\begin{equation}
\left\{e^M_N (5),\ \psi_M(8),\ A_M(3)\right\} 
\label{grav5}
\end{equation}
($M,N=1,\cdots,5$) where we have indicated in 
parenthesis the corresponding number of degrees of
freedom, the massless spectrum consists of $n_V=h_{(1,1)}-1$ vector multiplets 
\begin{equation}
\left\{A_{M}(3),\ \phi(1),\ \psi(4)\right\}
\label{vect5}
\end{equation}
and
$n_H=h_{(2,1)}+1$ hypermultiplets. The gauge group is abelian, $U(1)^{n_V+1}$,
where the additional factor counts the graviphoton.
Starting with the eleven-dimensional fields, 
\begin{equation}
\left\{e_I^J(44),\ A_{IJK}(84),\ \psi_I (128)\right\} 
\label{grav11}
\end{equation}
$(I,J,K=1,\cdots,11)$, and splitting the Lorentz 
indices as $(M,i,\bar\jmath)$ with 
$M=1,\dots,5$ and $i,\bar\jmath=1,2,3$, 
the $h_{(1,1)}$ gauge fields are given by 
$A_{M i\bar\jmath}$ while the $h_{(1,1)}-1$ vector moduli correspond to 
$g_{i\bar\jmath}$ with unit determinant. Moreover, the hypermultiplet moduli
are given by the $h_{(2,1)}$ complex scalar pairs 
$(g_{ij},A_{ij\overline{k}})$,
along with the universal scalars $(\det g_{i\bar\jmath},
A_{MNP},A_{ijk}=a\epsilon_{ijk})$.

The second step consists in the compactification of the previous 5D theory
down to four dimensions on $S^1/Z_2$, where the $Z_2$ acts as an inversion on
the fifth coordinate $y\rightarrow -y$ and changes the sign of the 11D 3-form
potential $A\to -A$~\cite{hw}. 
As a result, one obtains $N=1$ supersymmetry in four
dimensions together with $h_{(1,1)}+h_{(2,1)}+1$ massless chiral multiplets. The
corresponding scalar moduli are the $h_{(1,1)}$ real pairs $(g_{i\bar\jmath},
A_{5i\bar\jmath})$, the $h_{(2,1)}$ complex scalars $g_{ij}$ and the universal
real pair $(g_{55},A_{5\mu\nu})$. 

On top of the massless states, there is the
usual tower of their Kaluza--Klein excitations with masses
\begin{equation}
M^2={n^2\over\rho^2}\quad ;\quad n=0,\pm 1,\dots
\label{KK}
\end{equation}
corresponding to the fifth component of the momentum, $p_5$, which is 
quantized in units of the inverse radius of $S^1$, $1/\rho$. Because of the
$Z_2$ projection, only the symmetric combination of their excitations 
$|n\rangle +|$--$n\rangle$ survive. On the other hand, the $Z_2$-odd states
that were projected away at the massless level, give rise to massive
excitations corresponding to the antisymmetric combination 
$|n\rangle -|$--$n\rangle$. It follows that all states of the 5D theory appear 
at the massive level.

In addition to these untwisted fields, there
are twisted states located at the two end-points of the semicircle, giving
rise to the gauge group and to ordinary matter representations. 
In the case of the
standard embedding, there is an $E_8$ sitting at one of the end-points and an
$E_6$ with $h_{(1,1)}$ ${\bf 27}$ and $h_{(2,1)}$ $\overline{\bf 27}$
matter chiral multiplets sitting at the other. Of course, in any
realistic model, $E_6$ should be further broken down to the standard model
gauge group, for instance by turning on (discrete) Wilson lines.

\section{Supersymmetry breaking by Scherk-Schwarz 
on the eleventh dimension}
The $N=1$ supersymmetry transformations in the 5D theory are \cite{sierra}:
\begin{eqnarray}
\delta e_M^m &=& -{i\over 2}\overline{\cal E}\Gamma^m\Psi_M
\nonumber\\
\delta \Psi_M &=& D_M{\cal E} +\cdots
\label{susy}
\end{eqnarray}
where $e_M^m$ is the f\"unfbein, $\Gamma^m=(\gamma^\mu,i\gamma_5)$ are the
Dirac matrices, $\Psi_M$ is the gravitino field, ${\cal E}$ the
spinorial parameter of the transformation, and the dots stand for non-linear
terms. Similar transformations hold for the components of vector
multiplets and hypermultiplets for which our subsequent analysis 
can be generalized in a straightforward way. 

All fermions in eq.~(\ref{susy}) can be represented as doublets under the 
$SU(2)$ $R$-symmetry whose components are subject to the (generalized) Majorana
condition; in a suitable basis~\cite{cremmer}:
\begin{equation}
\Psi\equiv\left(
\begin{array}{c}\psi_1\\\psi_2\end{array}\right)
=\left(
\begin{array}{c}\gamma_5\psi_2^*\\-\gamma_5\psi_1^*\end{array}\right)\, ,
\label{doublet}
\end{equation}
where $\Psi$ describes any generic (Dirac) spinor.  It is
convenient to decompose the spinors with respect to the $\Gamma_5$
chirality. Using the relations $\gamma_5^2=1$ and 
$\gamma_5^*=-\gamma_5$, which are 
valid in the above basis, it follows that $\gamma_5\psi_1=\pm\psi_1$ implies 
$\gamma_5\psi_1^*=\mp\psi_1^*$. We can then define:
\begin{equation}
\Psi_{L,R}\equiv\left(
\begin{array}{c}\psi_{L,R}\\\pm\psi_{R;L}^*\end{array}\right);
\quad\Psi=\Psi_L+\Psi_R\, ,
\label{LR}
\end{equation}
in terms of the 4D chiral spinors $\psi_{L,R}=\pm\gamma_5\psi_{L,R}\, ,$ where
$\psi^*_{R,L}\equiv(\psi_{L,R})^*$. This decomposition amounts, 
in terms of $SU(2)_R$ doublets, to the condition
\begin{equation}
\Gamma_5\Psi_{L,R}=\pm\Psi_{L,R};\quad
\Gamma_5=\left(
\begin{array}{cc}\gamma_5&0\\0&-\gamma_5\end{array}\right)\ .
\label{chiral}
\end{equation}

We can now define the ${\cal R}$-chirality 
in terms of the spinors defined in eq.~(\ref{LR}), 
\begin{equation}
{\cal R}\Psi_{L,R}(x^\mu,y)=\pm\eta\Psi_{L,R}(x^\mu,-y)\ . 
\label{Rchiral}
\end{equation}
with $\eta=1$ for $\Psi=\Psi_\mu$ and $\eta=-1$ for
$\Psi=\Psi_5$. The 
${\bf Z}_2$ projection is defined by keeping the states that are even under
$\cal R$. It follows that the remaining massless fermions are the left-handed
components of the 4D gravitino $\Psi_{\mu L}$, as well as the
right-handed components of $\Psi_{5R}$. 
Taking into account the 
${\bf Z}_2$ action in the bosonic sector, which projects away the off-diagonal
components of the f\"unfbein ($e_\mu^5$), the above massless spectrum is 
consistent with the residual $N=1$ supersymmetry transformations at $D=4$
given by eq.~(\ref{susy}) with a fermionic parameter $\cal E$ reduced
to its left-handed component ${\cal E}_L$.

In order to spontaneously break supersymmetry,
we apply the Scherk-Schwarz mechanism on the fifth coordinate $y$ \cite{ss}. 
For this purpose, we need an $R$-symmetry, which transforms 
the gravitino non-trivially, and impose boundary conditions, around $S^1$, 
which are periodic up to a symmetry transformation:
\begin{equation}
\Psi_M(x^\mu,y+2\pi\rho)=e^{2i\pi\omega Q}\Psi_M(x^\mu,y)\, ,
\label{bc}
\end{equation}
where $Q$ is the $R$-symmetry generator and $\omega$ the transformation
parameter. The continuous symmetry is in general broken by the compactification
to some discrete subgroup, leading to quantized values of $\omega$. 
For instance,
in the case of ${\bf Z}_N$ one has $\omega=1/N$ and $Q=0,\dots,N-1$. For 
generic values of $\omega$,
eq.~(\ref{bc}) implies that the zero mode of the gravitino acquires
an explicit $y$-dependence:
\begin{equation}
\Psi_M(x^\mu,y)=U(y)\Psi^{(0)}_M(x^\mu)+\cdots;\;
{\displaystyle U=e^{i{\omega\over\rho}y Q}}
\label{mode}
\end{equation}
where the dots stand for Kaluza-Klein (KK) modes.

Consistency of the theory requires that the matrix $U$ commutes with the
reflection $\cal R$, which defines the $N=1$ 
projection~\cite{largeR,modelsR}. From eq.~(\ref{Rchiral}) one then finds:
\begin{equation}
\Gamma_5 U(-y)=U(y) \Gamma_5\, ,
\label{cons}
\end{equation}
implying that the generator $Q$ anticommutes with $\Gamma_5$, 
$\{Q,\Gamma_5\}=0$. Notice that condition (\ref{cons}) guarantees 
that the $\cal R$-chirality of a spinor, $\Psi_{L,R}(x^\mu,y)$, 
in the sense of eq.~(\ref{Rchiral}), coincides
with the $\Gamma_5$-chirality of its zero-mode $\Psi_{L,R}^{(0)}(x^\mu)$, in 
the sense of eq.~(\ref{chiral}). In this way one can write the decomposition
(\ref{mode}) for the chiral components of $\Psi$, i.e.
$\Psi_{L,R}(x^\mu,y)=U(y)\Psi_{L,R}^{(0)}(x^\mu)$.
A solution is given by \cite{dg} (for the general solution see \cite{m2}):
\begin{equation}
Q=\sigma_2;\quad 
U=\cos{\omega y\over\rho}+i\sigma_2\sin{\omega y\over\rho}\, ,
\label{Umatrix}
\end{equation}
where $\sigma_{i}$ are the Pauli matrices representing 
$SU(2)_R$ generators. 

For the particular value $\omega=1/2$ there is an
additional solution to eq.~(\ref{cons}) \cite{m2},
\begin{equation}
Q=1\quad ;\quad U=-\cos\frac{y}{2\rho}\ ,
\label{F}
\end{equation}
which, acting on the 5D fields, consists on 
$\exp\{i\pi Q\}=(-)^{2s}$, changing the sign of fermions 
and leaving bosons invariant. This solution involves both $n=0$ and
$n=-1$ KK-modes, which makes the effective field theory description of the
spontaneous supersymmetry breaking more complicated. For this reason we restrict
the analysis in this section to the solution (\ref{Umatrix}).

Inspection of the supersymmetry transformations (\ref{susy}), together with 
the requirement that the f\"unfbein zero mode does not depend on $y$, 
$e^m_M=e^m_M(x^{\mu})$, shows
that the $y$-dependence of the supersymmetry parameter is the same as that of
the gravitino zero-mode \cite{ss}, i.e. 
\begin{equation}
{\cal E}(x^{\mu},y)=U(y){\cal E}^{(0)}(x^{\mu})\ .
\label{sstransf}
\end{equation}
Supersymmetry in the 4D theory is then spontaneously broken, with 
the goldstino being identified with the fifth component of the 5D gravitino,
$\Psi_5^{(0)}$. Indeed, for global supersymmetry parameter,
$D_\mu{\cal E}^{(0)}=0$, its variation is [Note that the operator
$U^{-1}\partial_y U$ turns a left-handed spinor, in the sense of
eq.~(\ref{chiral}), into a right-handed one.]:
\begin{equation}
\delta\Psi_5^{(0)}=i\sigma_2{\omega\over\rho}{\cal E}^{(0)} +\cdots
\label{gold}
\end{equation}
while no other fermions can acquire finite constant shifts in their 
transformations.

The above arguments are also valid in the $N=1$ theory, 
obtained by applying the
${\bf Z}_2$ projection defined through the $\cal R$-reflection (\ref{Rchiral}).
The $y$-dependence of the remaining zero modes is always given by
eq.~(\ref{mode}). Supersymmetry is spontaneously broken:
\begin{itemize}
\item
The goldstino is identified as the right-handed component 
\begin{equation}
{\rm goldstino}\equiv\psi^{(0)}_{5R}, 
\label{goldstino}
\end{equation}
which, from eq.~(\ref{gold}), transforms as:
\begin{equation}
\delta\psi_{5R}^{(0)}={\omega\over\rho}\varepsilon^{(0)*}_R+\cdots
\label{goldR}
\end{equation}
\item
The surviving gravitino is $\Psi^{(0)}_{\mu L}$ in the notation of
eq.~(\ref{LR}). Its mass is given by
\begin{equation}
m_{3/2}=\frac{\omega}{\rho}
\label{gravmasa}
\end{equation}
\item
In the limit $\rho\rightarrow\infty$, supersymmetry is locally restored:
$m_{3/2}\rightarrow 0$, $\delta\psi_{5R}^{(0)}\rightarrow 0$.
\end{itemize} 

Note, however, that the above analysis in the $N=1$ case is
valid, strictly speaking, for values of
$y$ inside the semicircle, obtained from the interval $[-\pi\rho,\pi\rho]$ 
through the identification $y\leftrightarrow -y$. This leads to a discontinuity
in the transformation parameter $\cal E$
around the end-point $y=\pm\pi\rho$, since 
$U(-\pi\rho)=U^{-1}(\pi\rho)$:
\begin{equation}
{\cal E}(-\pi\rho)\neq{\cal E}(\pi\rho)\, .
\label{neq}
\end{equation}
This discontinuity survives even in the large-radius limit $\rho\to\infty$ 
where the gravitino mass vanishes and supersymmetry is restored locally. This
phenomenon is reminiscent of the one found in ref.~\cite{h}, where the
discontinuity at the weakly coupled end $y=\pi\rho$ is due to the gaugino
condensate of the hidden $E_8$ formed at the strongly coupled end $y=0$. 
In fact
the two results become identical for the transformation parameter ${\cal E}$ in
the neighbourhood $y\sim\pi\rho$, in the limit $\rho\to\infty$:
\begin{equation}
\varepsilon_L(y)\sim
\cos\pi\omega\, \varepsilon^{(0)}_L+
\epsilon(y)\sin\pi\omega\,\varepsilon^{(0)*}_R\, .
\label{lim}
\end{equation}
On the other hand, it is easy to see that the goldstino variation vanishes in
this limit, since the discontinuity in $\partial_y\varepsilon(y)_L$ is
proportional to $\delta(y)\sin(y\omega/\rho)$.
The transformation parameter $\varepsilon_L(y)$ is thus identified 
with the spinor $\eta'$ of ref.~\cite{h}, which
solves the unbroken supersymmetry condition $\delta\psi_{5R}=0$.

\section{Supersymmetry breaking in the observable sector}

At the lowest order, supersymmetry is broken only in the five-dimensional 
bulk (gravitational and moduli sector), while it remains
unbroken in the observable sector. The communication of supersymmetry breaking
is then expected to arise radiatively, by gravitational interactions. This
issue is studied below in the particular case $\omega=1/2$ and 
$\exp\{i\pi Q\}=(-)^{2s}$ \cite{m}, though more general cases can be equally 
computed \cite{preparation}.

\subsection{Scalar masses}

At the one-loop level, the diagrams that contribute to the scalar masses in the
observable sector were studied in ref.~\cite{m}, 
where the vertices come from the kinetic terms. 
\begin{equation}
{\cal G}^{\varphi\varphi}_{z^{(n)}}\, \phi \overline{\psi}_{\varphi}
\gamma_{\mu} D^{\mu}\chi^{(n)}+\cdots \ .
\label{kin}
\end{equation}
{}Fields from the boundary, generically denoted by $(\varphi,\psi_\varphi)$, 
always appear in pairs, 
as dictated by the $Z_2$ invariance. Moreover, in the effective field 
theory limit $\rho M_{11}\gg 1$, their couplings to fields from the bulk 
are the same for all Kaluza--Klein excitations. The latter are the moduli
$(z^{(n)},\chi^{(n)})$ and the graviton $(g_{\mu\nu}^{(n)},\psi_\mu^{(n)})$ 
supermultiplets. Moduli and matter field indices are dropped for notational
simplicity.

After adding the contribution of diagrams related by supersymmetry, 
we obtain the following expression for the scalar masses:
\begin{equation}
m_{\varphi{\bar\varphi}}^2\sim G^{-1}_{\varphi{\bar\varphi}}\,
\left( G^{i{\bar\jmath}}R_{i{\bar\jmath}\varphi{\bar\varphi}}-
G_{\varphi{\bar\varphi}}\right)
{m_{3/2}^4\over M_p^2}{\cal J}\ ,
\label{mphi}
\end{equation}
where we used the relation (\ref{gravmasa}) for $m_{3/2}$, 
and ${\cal J}$ is a constant given by:
\begin{equation}
{\cal J}=\int_0^\infty{dx\over x^3}\left({\pi\over x}\right)^{1/2}
\left[ \theta_3\left({i\pi\over x}\right)-
\theta_4\left({i\pi\over x}\right)\right]
\label{Jintegral}
\end{equation}
where $\theta_i$ are the Jacobi theta-functions and 
we have used the Poisson resummation formula. In eq.~(\ref{mphi}),
$G_{i{\bar\jmath}}$ and $G_{\varphi{\bar\varphi}}$ are the moduli and matter 
metrics, while 
$R_{i{\bar\jmath}\varphi{\bar\varphi}}$ is the moduli-matter mixed Riemann
tensor. The factor $G^{-1}_{\varphi{\bar\varphi}}$ comes from the 
wave function renormalization and
the two terms in the bracket correspond to the contributions 
of the moduli and graviton supermultiplets. 
As a result, we find the scalar masses 
$m_{\varphi{\bar\varphi}}={\cal O}(10^{-1})\; m_{3/2}^2/M_p\sim 10^4$~GeV 
generically. This is only a rough estimate, since besides the moduli 
dependent prefactor in eq.~(\ref{mphi}), the result is very sensitive to the 
value of $M_{11}$. In fact, eqs.~(\ref{M11}) and (\ref{gravmasa}) show that 
$m_{\varphi{\bar\varphi}}$ scales as
$M_{11}^6$ and, thus, a modest factor of 2 in $M_{11}$ changes the scalar
masses by
almost two orders of magnitude.

A similar analysis can be applied to compute the masses of the scalar moduli.
The evaluation of the corresponding diagrams yields \cite{m}:
\begin{equation}
m_{z{\bar z}}^2\sim 5\, G^{-1}_{z{\bar z}}
\left( R_{z{\bar z}}-G_{z{\bar z}}\right)
{m_{3/2}^4\over M_p^2}{\cal J}\ ,
\label{mz}
\end{equation}
where $R_{z{\bar z}}$ is the moduli Ricci tensor and
the constant ${\cal J}$ is given in eq.~(\ref{Jintegral}). Thus, all moduli
scalars obtain masses of the same order as the scalar masses in the observable
sector, ${\cal O}(10)$ TeV.

The fact that all scalar squared mass splittings are of order $m_{3/2}^4/M_p^2$
is a consequence of the absence of quadratic divergences in the effective
supergravity. Inspection of eq.~(\ref{Jintegral}) shows that cancellation of 
quadratic divergences arises non-trivially. Indeed, any single
excitation $n$ of the sum gives a contribution to the integral, which is
quadratically divergent at $x=0$ as $dx/x^2$, so that after introducing an
ultraviolet cutoff $\propto 1/M_p^2$ one would get a contribution of
order $m_{3/2}$ to the mass. 
However, after summing over all modes and performing the
Poisson resummation, one finds that the integrand has an exponentially 
suppressed (non-analytic) ultraviolet behaviour as 
$e^{-\pi^2/x}/x^{5/2}$.
One can also compute the effective potential as a function 
of the background $\rho$:
\begin{equation}
V_{\rm eff}=-{N{\cal J}\over 32\pi^2}{1\over\rho^4}\ ,
\label{Veff}
\end{equation}
where $N$ is the number of massless multiplets from the bulk. This result
explicitly shows the vanishing of Str$M^2$ after supersymmetry breaking.

\subsection{Gaugino masses}

Gaugino masses also receive radiative gravitational contributions. 
At the one-loop level they lead to individual contributions:
\begin{equation}
m_{\lambda\lambda}\propto{m_{3/2}^3\over M_p^2}\ ,
\label{mg}
\end{equation}
where we followed the same steps as in the case of scalar masses.

The above result shows that the one-loop gravitational contributions to 
gaugino masses are too small for phenomenological purposes. This is a general
problem, which has been known for a long time~\cite{gcold1,gcold2}. 
A possible solution exists if there are massive
matter fields transforming non-trivially under the gauge group. Then, their
mass splittings generate gaugino masses by one-loop diagrams involving gauge
interactions. The latter lead to finite contributions given by~\cite{bgm}:
\begin{equation}
m_{\lambda\lambda}\sim N_s
{\alpha\over 2\pi}\mu f\left({m_s\over\mu}\right)\ ,
\label{f}
\end{equation}
where $\mu$ is the supersymmetric mass and $m_s^2$ the squared mass 
splitting of those matter fields; $N_s$ denotes their multiplicity,
$\alpha$ is the corresponding gauge coupling, 
and the function $f(x)$ is nearly constant for $x\simgt 2$ 
while it behaves as $x$ for $x\simlt 1$. Thus, the gaugino masses are
of the order of $(\alpha/2\pi)N_s\, {\rm min}(\mu,m_s)$.

It is easy to see that when $\mu$ is below the intermediate scale $\rho^{-1}$, 
the evaluation of the scalar masses (\ref{mphi}), (\ref{Jintegral})
remains valid up to ${\cal O}(\mu/m_{3/2})$
corrections. It follows that the gaugino masses are approximately one order of
magnitude lower than the scalar masses if $\mu\simgt m_s$. Although 
this mechanism can give acceptable masses to charginos and neutralinos,
provided that the Higgs supersymmetric parameter $\mu$ is large enough, gluino
masses would require the Standard Model particle content to be extended by the
presence of extra colour multiplets in vector-like
representations such as triplets or leptoquarks. 
Of course, in this case, unification requires that the extra matter appears in
complete $SU(5)$ representations, e.g. $({\bf 5}+\overline{\bf 5})$ or 
$({\bf 10}+\overline{\bf 10})$.
Otherwise, in the absence of extra matter, this scenario
leaves open the possibility of having light gluinos with masses of order
$(\alpha_3/2\pi)\, m_{\rm top}={\cal O}(1)$ GeV \cite{farrar}.

To summarize, the mass spectrum we obtained in the observable sector 
originates from local supersymmetry breaking, with a gravitino mass $m_{3/2}$ 
at an intermediate scale defined by the size of the eleventh dimension of 
M-theory. All scalars then acquire masses of order 
$m_{3/2}^2/M_p$, while gaugino masses are of order $m_{3/2}^3/M_p^2$.
This situation is again identical to the case where supersymmetry is broken in
the heterotic string by gaugino condensation stabilized by a VEV of the 
antisymmetric tensor field strength~\cite{gcold1,gcold2}. 
As we saw, the problem of having very
light gauginos can be solved by means of gauge interactions involving extra
{}fields and providing gaugino masses of order $(\alpha/2\pi)m_{3/2}^2/M_p$.
Therefore, this scenario predicts a hierarchy of supersymmetric mass spectrum
where scalars are much heavier than gauginos.

{}Finally, in the old analysis of gaugino condensation, 
based on the heterotic string tree-level effective supergravity, 
it was found that scalars remained massless at the one-loop order,
because of the dilation properties of the K{\"a}hler potential~\cite{gcold1}. 
In fact, it is easy
to see that the term in the brackets of eq.~(\ref{mphi}) vanishes when the
K{\"a}hler potential has for instance the no-scale $SU(1,n)$ form 
$K=-3\ln (z+{\bar z}- |\varphi|^2)$ and $\varphi$'s have zero VEVs.
This can lead to an alternative scenario where gauginos, with masses of order 
$m_{3/2}^3/M_p^2\sim 1$ TeV (for $m_{3/2}\sim 10^{14}$ GeV),
seed supersymmetry breaking in the rest of
the observable sector by gauge interactions. Since in this case the
corresponding diagrams are logarithmically divergent, all supersymmetric
masses turn out to be of the same order of magnitude. However, this scenario
is expected (and was explicitly shown~\cite{gcold2}) to be unstable 
under higher-order loop corrections, 
as the dilation symmetry is in general broken at the quantum level.

\section{Relation with gaugino condensation}

Supersymmetry breaking by Scherk-Schwarz compactification of
the eleventh dimension reproduces the main features (at least in
the simplest case) of gaugino condensation in the weakly coupled
heterotic string. Then, it is natural to conjecture that it provides a
dual description of gaugino condensation in the strongly coupled
regime \cite{m,m2}.

\subsection{The weakly coupled heterotic string}

On the heterotic side, one expects that (local) supersymmetry can be broken by
gaugino condensation effects in the hidden $E_8$, at least in the (10D) weakly
coupled regime~\cite{gc1}. Let us briefly describe
the main features of this
mechanism. The physical picture is that the condensate
$\langle\lambda\lambda\rangle$ develops at a scale $\Lambda_c$, where the gauge
coupling of the hidden $E_8$ becomes strong:
\begin{equation}
{\displaystyle
\langle\lambda\lambda\rangle\sim\Lambda_c^3=
\mu^3e^{-{2\pi\over c_8\alpha_8(\mu)}}}\ ,
\label{ll}
\end{equation}
with $c_8=30$ being the quadratic Casimir of $E_8$ and $\alpha_8(\mu)$ 
its coupling constant at the scale $\mu$.

This phenomenon can be described by introducing a chiral supermultiplet $U$
whose vacuum expectation value (VEV) reproduces the condensate 
(\ref{ll})~\cite{gc2}. The
effective non-perturbative superpotential is determined by consideration of the
anomalous Ward identities:
\begin{equation}
W_{\rm np}\propto U\left(
{1\over\alpha_W}+{c_8\over2\pi}\ln{U\over \mu^3}\right)\ ,
\label{w}
\end{equation}
where $\alpha_W$ is the Wilsonian effective coupling (at the scale $\mu$), 
which depends holomorphically on the moduli~\cite{sv}. 
It is related to the physical coupling by:
\begin{equation}
{1\over\alpha_8}={\rm Re}{1\over\alpha_W}+{c_8\over 4\pi}
(-K+2\ln(S+{\bar S}))\ ,
\label{awa8}
\end{equation}
where $K$ is the K{\"a}hler potential and $S$ is the heterotic dilaton whose
VEV determines the 4D string coupling constant, Re$S=1/\alpha_G$.

Minimization of the effective potential 
with respect to $U$ implies to leading order in $\Lambda_c/M_p$ the condition 
$\partial_U W_{\rm np}=0$~\cite{bdqq}, which gives
\begin{equation}
{\displaystyle
U=\mu^3e^{-{2\pi\over c_8\alpha_W}-1}\qquad ;\qquad
W_{\rm np}\propto U}\ .
\label{U}
\end{equation}
Using this result together with eqs.~(\ref{ll}) and (\ref{awa8}), it is
straightforward to obtain the value of the gravitino mass:
\begin{equation}
m_{3/2}=|W_{\rm np}|e^{K/2}\propto {1\over\alpha_G}\Lambda_c^3M_p^{-2}\ .
\label{m32}
\end{equation}

The effective potential should also be minimized with respect to the dilaton
{}field $S$. Unfortunately, its runaway behaviour drives the theory to the
supersymmetric limit with vanishing coupling, $S\rightarrow\infty$. A
possible stabilization mechanism was initially proposed by means of a
VEV for the field-strength of the antisymmetric tensor field along the compact
directions, which shifts the superpotential by a constant~\cite{gc1}. However,
this constant was found to be quantized, so that $W_{\rm np}$ becomes of order
one at the minimum~\cite{rw}. Then, eq.~(\ref{m32}) implies that the only way
to obtain a hierarchy for the gravitino mass is by making $e^{K/2}$ small, or
equivalently by having a large compactification volume $V\sim e^{-K}$.
As a result, we obtain the following scaling relations (in $M_p$ units):
\begin{equation}
m_{3/2}\sim V^{-1/2}; \quad \Lambda_c\sim V^{-1/6}\ .
\label{scaling}
\end{equation}

Assuming that eqs.~(\ref{m32}) and (\ref{scaling}) hold in the strong 
coupling regime, a comparison with the duality relations (\ref{M11})
implies the identification of
the condensation scale $\Lambda_c$ with the M-theory scale $M_{11}$
and the inverse radius of the semicircle $\rho^{-1}$ with the gravitino mass:
\begin{equation}
\Lambda_c\sim M_{11}\quad ; \qquad m_{3/2}\sim \rho^{-1}\ .
\end{equation}

\subsection{M-theory}

In the strongly coupled regime:

\begin{itemize}
\item
The relation $m_{3/2}\sim \rho^{-1}$ is provided by the
Scherk-Schwarz mechanism as we have seen in previous sections. 
\item
In the description of gaugino condensation by the Scherk-Schwarz mechanism, the
condensation scale is identified with the M-theory scale $M_{11}$. This implies
that the hidden $E_8$ is strongly coupled and should not contain any massless 
matter in the perturbative spectrum. Consistency then requires that the
corresponding gauge coupling be large, $\alpha_8(M_{11})\simgt 1$. 
\end{itemize}
On the M-theory side
this provides a constraint that naively fixes the 4D unification coupling
$\alpha_G$ to be in a non-perturbative regime. Fortunately, there are important
M-theory threshold effects that invalidate this conclusion. These effects can
be understood from the lack of factorization of 
the 7-dimensional internal space
as a direct product of the semicircle with a Calabi-Yau manifold, CY$\times
S^1/{\bf Z}_2$ \cite{w}. As a result, the Calabi-Yau volume $V$ 
becomes a function
of $\rho$ and takes different values at the two end-points of the semicircle. 
In the large-radius limit, one finds \cite{w}:
\begin{eqnarray} & & 
V(0)=V(\pi\rho)-{1\over 32\pi^2}\rho M_{11}^{-3}
\label{thr}
\\ & &
\left|\int_{\rm CY}{\omega\over 4\pi^2}\wedge
\left({\rm tr} F'\wedge F'-{\rm tr} F\wedge F\right)\right|\, ,
\nonumber
\end{eqnarray}
where $\omega$ is the K\"ahler form of CY and $F'$ $(F)$ is the field 
strength of the strongly
(weakly) coupled $E_8$ sitting at the end-point $y=0$ $(y=\pi\rho)$. 
The integral in the r.h.s. is 
a linear function of the $h_{(1,1)}-1$ K\"ahler class moduli for unit
volume, which belong to 5D vector multiplets. Its natural value is
$M_{11}^{-2}$ up to a proportionality factor of order~1~\cite{w}.

Following eq.~(\ref{M11}), the gauge coupling constants at the two 
end-points are related to the corresponding volumes as \cite{hw}:
\begin{equation}
{1\over\alpha_G}=2M_{11}^6V(\pi\rho);\;
{1\over\alpha_8(M_{11})}=2M_{11}^6V(0)\, ,
\label{alphas}
\end{equation}
Imposing now the constraint 
\begin{equation}
\alpha_8(M_{11})\simgt 1;\quad M_{11}\sim \Lambda_c
\label{alpha8}
\end{equation}
and using eqs.~(\ref{thr}) and
(\ref{alphas}), one finds $\rho\sim\rho_{\rm crit}$ where $\rho_{\rm
crit}$
corresponds to the critical value at which the volume at the strongly coupled
end vanishes and the hidden $E_8$ decouples from the low-energy spectrum:
\begin{equation}
\rho_{\rm crit}^{-1}\sim{\alpha_G\over 16\pi^2}\, 
M_{11}\simeq 2\times 10^{-4}M_{11}
\label{crit}
\end{equation}
Note that this condition can also be thought of as resulting from 
a minimization of 
the (positive semi-definite) 4D gaugino condensation potential, which is 
proportional to $V(0)$ and, thus, vanishes at zero volume.
It is remarkable that the above relation provides the hierarchy 
necessary to fix
$\rho^{-1}$ at the intermediate scale $\sim 10^{12}$ GeV, when one identifies
the M-theory scale $M_{11}$ with the unification mass $\sim 10^{16}$ GeV
inferred by the low-energy data \cite{w,m0}.

\begin{itemize}
\item
The $\rho\rightarrow\infty$ limit
\end{itemize}

We have already mentioned that in the $\rho\rightarrow\infty$
limit both gaugino condensation and Scherk-Schwarz mechanisms
lead to similar conclusions on supersymmetry breaking. In fact
the proportionality constant $\sin\pi\omega$ 
in eq.~(\ref{lim}) plays the role of the
gaugino condensate in the dual description and vanishes only for
integer values of $\omega$ for which the Scherk-Schwarz mechanism becomes
trivial. In general $\omega$ is quantized, as we discussed earlier, which is
consistent with the quantization of the 
gaugino condensate through its equation of
motion that relates it with the VEV of the antisymmetric tensor field 
strength \cite{rw}. 

In the presence of gaugino condensation, the discontinuity 
in the function $\epsilon(y)$ (\ref{neq})
was interpreted as a topological obstruction 
that signals supersymmetry breaking when effects of finite radius
would be taken into account \cite{h}. 
Here we have shown that the same discontinuity,
in the infinite-radius limit, is reproduced by the Scherk-Schwarz mechanism.
 
\section{Conclusion}
$\bullet$ The eleventh dimension of the M-theory seems an
interesting candidate to perturbatively break supersymmetry in
the gravitational and moduli sector.

$\bullet$ The Scherk-Schwarz mechanism of M-theory is not a
single model but a framework where many different models can be
accomodated. If for instance we use (a $U(1)$ subgroup of) the
$SU(2)_R$, then all fermions of the vector
multiplets and all complex scalars of the hypermultiplets transform 
in a similar fashion as the gravitino \cite{dg}.
However, if we use $(-)^{2s}$ ($\omega=1/2$), it is acting non-trivially 
only on the fermions of both vector multiplets and hypermultiplets~\cite{m}.

$\bullet$ This mechanism provides an alternative ``perturbative" explanation 
of the gauge
hierarchy, where the smallness of the ratio $m_{\rm susy}/M_p\sim 10^{-16}$ 
is provided by powers of the unification coupling $\sim(\alpha_G/16\pi^2)^4$
instead of the conventional non-perturbative suppression $\sim e^{-1/\alpha_G}$.
Of course in both cases, the remaining open problem is to determine 
the actual value of the gauge coupling $\alpha_G$. 
In the present context of M-theory, this amounts to 
fixing the volume of the Calabi-Yau manifold $V(\pi\rho)$. 

$\bullet$ The features of supersymmetry breaking by the
Scherk-Schwarz mechanism are similar to (some) models of
non-perturbative supersymmetry breaking by gaugino condensation
in the weakly coupled heterotic string.

$\bullet$ One of the main open problems is to find the general features 
of the low-energy effective theory describing the mechanism of
supersymmetry breaking, 
and the proposed equivalence
between the {\it perturbative} breaking of supersymmetry in the
M-theory, by the Scherk-Schwarz mechanism on the eleventh
dimension~\cite{dg}, and the {\it non-perturbative} breaking by gaugino
condensation in the heterotic string~\cite{nilles,lalak}.

\end{document}